\documentstyle[twocolumn,aps]{revtex}
\begin{document}
\input {psfig.sty}
\draft

%
\twocolumn[\hsize\textwidth\columnwidth\hsize\csname@twocolumnfalse\endcsname
%
%

\title{Effect of Adiabatic Phonons on Striped and Homogeneous Ground States}

\author{Yucel Yildirim$^{1,2}$ and Adriana Moreo$^{1}$}

\address{$^1$Department of Physics and Astronomy,University of Tennessee,
Knoxville, TN 37966-1200 and 
\\Oak Ridge National Laboratory,Oak Ridge, 
TN 37831-6032 }

\address{$^2$Department of Physicsand  National High Magnetic Field Lab,
\\ Florida State University, Tallahassee, FL 32306, USA}

\date{\today}
\maketitle

\begin{abstract}

The effects of adiabatic phonons on a spin-fermion model for high $T_c$ 
cuprates are studied using numerical simulations. In the absence of 
electron-phonon interactions (EPI),
stripes in the ground state are observed \cite{adri1} for 
certain dopings while homogeneous states are stabilized in other regions 
of parameter space.
Different modes of adiabatic phonons are added to the Hamiltonian:
breathing, shear and half-breathing modes. Diagonal and 
off-diagonal electron-phonon couplings are
considered. It is observed that strong diagonal EPI generate stripes in 
previously homogeneous states, while in striped ground states
an increase in the diagonal couplings
tends to stabilize the stripes, inducing a gap in the density of states (DOS)
and rendering the ground state insulating. The off-diagonal terms, on the 
other hand,  
destabilize the stripes creating inhomogeneous ground states with a pseudogap 
at the chemical potential in the DOS. The breathing mode stabilizes static 
diagonal stripes; while the half-breathing (shear) modes 
stabilize dynamical (localized) vertical and horizontal stripes. The EPI 
induces decoherence of the quasi-particle peaks in the spectral functions.

\end{abstract}

\vskip2pc]
\narrowtext

\section{Introduction}

The pairing mechanism responsible for high $T_c$ superconductivity is still 
unknown. The electron-phonon interactions that satisfactorily explain pairing 
for traditional superconductors within the BCS theory \cite{BCS} would 
require phonon frequencies incompatible with the material stability in order 
to produce the observed high critical temperatures in the 
cuprates.\cite{BCS}  
For this reason many researchers believe that magnetic interactions, 
which are observed in all the cuprates, may play an important role in the 
pairing mechanism.\cite{magnetic} As a result of this hypothesis, most of the 
Hamiltonians proposed to study the physics of the cuprates, 
such as the Hubbard and t-J models, only 
incorporate electronic and magnetic degrees 
of freedom.\cite {review} However, experiments indicate that there are 
active phonon modes in the cuprates.\cite{Bianconi,Egami,Lanzara,Tranquada} 
In addition, experimentally a very rich phase diagram, particularly in the 
underdoped regime, is started to be unveiled. Some materials appear to have 
ground states with electronic stripes as observed in neutron scattering results
\cite{Tranquada}, while scanning tunneling microscopy indicates nanosize 
patches of superconducting and non-superconducting phases.
\cite{Davis,Yazdani} The emerging phase complexity is reminiscent of the 
experimental data for manganites where competing electronic, magnetic and 
phononic degrees of freedom are responsible for the rich phase 
structure.\cite{manrev} 

For these reasons it is important to 
include electron-phonon interactions in models for the cuprates. 
This would allow to understand whether EPI stabilize or destabilize charge 
stripes, what kind of inhomogeneous textures, if any, develop and,
eventually, whether the interplay of magnetic and phonon interactions with 
the electrons is responsible for the pairing mechanism.\cite{bili}

The first step towards the goal of introducing EPI in models for the cuprates 
is to propose a simple but physically realistic Hamiltonian that can be 
studied with unbiased techniques. The proposals already in the literature
include momentum dependent electron-phonon couplings that lead to long-range 
interactions in coordinate space\cite {Olle,Dev} 
and/or quantum phonons which are 
very difficult to treat numerically.\cite{jose,doug} 
Most studies have been performed 
using mean-field, slave-boson, or LDA approximations.\cite{bishop,ishihara}
Numerical simulations have been done using the $t-J$ model in very small
lattices, with a limited number of phonon modes and diagonal couplings,
\cite{jose,doug} or on the one-dimensional Hubbard model.\cite{tachi}

In this paper we will study numerically a spin-fermion (SF) 
Hamiltonian for the 
cuprates\cite{adri1} with electron-phonon interactions. This model reproduces 
many properties of the cuprates and presents stripes in the ground state, 
due solely to spin-charge interactions, 
in some regions of parameter space.\cite{adri1} 
Thus, it provides a framework particularly suitable to study the 
effects of electron-phonon interactions on the preformed stripes. However, 
charge homogeneous ground states are also found in other regions of parameter 
space which allows to investigate charge inhomogeneity induced by EPI. 
Several phononic modes 
will be studied and diagonal and non-diagonal couplings, i.e., the 
dependence of the hopping and other Hamiltonian parameters on the lattice 
distortions, will be considered. The work will be performed in the adiabatic 
limit, i.e., at zero phononic frequency.      
 
The paper is organized as follows: in Section II the Hamiltonian is 
introduced; the effect of electron-phonon interactions on striped states 
are presented in Section III, while Section IV is devoted to the effects 
of EPI on homogeneous states.  Section V contains the Conclusions. 

\section {The model}

The SF-model is constructed as an interacting system of
electrons and spins, mimicking phenomenologically the
coexistence of charge and spin degrees of freedom in 
the cuprates \cite{Pines}. Its Hamiltonian is given by
$$
{\rm H=
-t{ \sum_{\langle {\bf ij} \rangle\alpha}(c^{\dagger}_{{\bf i}\alpha}
c_{{\bf j}\alpha}+h.c.)}}
+{\rm J
\sum_{{\bf i}}
{\bf{s}}_{\bf i}\cdot{\bf{S}}_{\bf i}
+J'\sum_{\langle {\bf ij} \rangle}{\bf{S}}_{\bf i} \cdot{\bf{S}}_{\bf j}},
\eqno(1)
$$
\noindent where ${\rm c^{\dagger}_{{\bf i}\alpha} }$ creates an electron
at site ${\bf i}=({\rm i_x,i_y})$ with spin projection $\alpha$,  
${\bf s_i}$=$\rm \sum_{\alpha\beta} 
c^{\dagger}_{{\bf i}\alpha}{\bf{\sigma}}_{\alpha\beta}c_{{\bf
i}\beta}$ is the spin of the mobile electron, the  Pauli
matrices are denoted by ${\bf{\sigma}}$,
${\bf{S}_i}$ is the localized
spin at site ${\bf i}$,
${ \langle {\bf ij} \rangle }$ denotes nearest-neighbor (NN)
lattice sites,
${\rm t}$ is the NN-hopping amplitude for the electrons,
${\rm J>0}$ is an antiferromagnetic (AF) coupling between the spins of
the mobile and localized degrees of freedom,
and ${\rm J'>0}$ is a direct AF coupling
between the localized spins.
The density $\rm \langle n \rangle$=$\rm 1-x$ of 
itinerant electrons is controlled by a chemical potential $\mu$. 
Hereafter ${\rm t=1}$ will be used as the unit of energy. 
${\rm J'}$ and ${\rm J}$ are fixed to 0.05 and 2.0 respectively, 
values shown to be realistic in previous investigations \cite{adri1}. 
The temperature will be fixed to a low value:
T=0.01, which was shown before to lead 
to the correct high-$T_c$ phenomenology. \cite{adri1,moham}

The diagonal electron-phonon part of the Hamiltonian being proposed here 
is given by
$$
H_{\rm e-ph}^{(j)}=-\lambda\sum_{{\bf i}}Q_{{\bf i}}^{(j)}n_{{\bf i}},
\eqno(2)
$$
\noindent where $n_{{\bf i}}=\sum_{\sigma}c^{\dagger}_{{\bf i}\sigma}
c_{{\bf i}\sigma}$ is the electronic density on site ${\bf i}$ and
$Q_{{\bf i}}^{(j)}$ is the phonon mode defined in terms of the lattice 
distortions $u_{{\bf i},\alpha}$ which measures the displacement along the
directions $\alpha=\hat x$ or $\hat y$ of oxygen ions located at the center 
of the lattice's links in the equilibrium position, i.e., 
$u_{{\bf i},\alpha}=0$. The index $(j)$ identifies the phonon mode. In this 
work, the following phonon modes will be considered: 

\noindent (a) The {\it breathing} mode
given by
$$
Q_{{\bf i}}^{(1)}=\sum_{\alpha}( u_{{\bf i},\alpha}-u_{{\bf i-\hat\alpha},\alpha});
\eqno(3)
$$
\noindent (b) The {\it shear} 
mode, in which the oxygens along e.g., $x$ move in counterphase 
with the oxygens along $y$, given by
$$
Q_{{\bf i}}^{(2)}=\sum_{\alpha}(-1)^{\sigma}
( u_{{\bf i},\alpha}-u_{{\bf i-\hat\alpha},\alpha}),
\eqno(4)
$$
\noindent with $\sigma=1(-1)$ for $\alpha= x(y)$; 

\noindent (c) The {\it  half-breathing}
mode along $x$ given by
$$
Q_{{\bf i}}^{(3)}=( u_{{\bf i}, x}-u_{{\bf i-\hat x}, x});
\eqno(5)
$$
\noindent and (d) The {\it half-breathing} mode along $y$ given by
$$
Q_{{\bf i}}^{(4)}=( u_{{\bf i}, y}-u_{{\bf i-\hat y}, y}).
\eqno(6)
$$

Note that although the proposed interactions seem local in coordinate space,
they correspond to cooperative lattice distortions which, in turn, will 
produce strongly momentum dependent effective electron-phonon couplings, 
in agreement with the experimental evidence observed in the cuprates.
\cite{Egami}

A term to incorporate the stiffness of the Cu-O bonds is added. The term 
bounds the amplitude of  
the lattice distortions induced by $H_{\rm e-ph}$. Its explicit form is:
$$
H_{\rm ph}=\kappa\sum_{{\bf i},\alpha}(u_{{\bf i},\alpha})^2,
\eqno(7)
$$
\noindent where $\kappa$ is the stiffness parameter that will be set to 1 
here. In addition, we will 
consider the off-diagonal interactions induced by the lattice distortions.
To obtain these terms we follow the approach of Ishihara {\it et al}.
\cite{tachi,ishihara} As a result the hopping $t$ in Eq.(1) now becomes site 
and direction dependent and it is given by
$$
t_{{\bf i,j}}=t+\gamma[u({\bf i})+u({\bf j})],
\eqno(8)
$$
\noindent where $\gamma$ is a parameter and 
$$
u({\bf i})=u_{{\bf i}, x}-u_{{\bf i-\hat x}, x}+
u_{{\bf i}, y}-u_{{\bf i-\hat y}, y}.
\eqno(9)
$$
The Heisenberg coupling $J'$ in Eq.(1) also is affected by the lattice 
distortions and it has to be replaced by
$$
J'_{{\bf i,j}}=J'+g_J\gamma[u({\bf i})+u({\bf j})],
\eqno(10)
$$
\noindent where $g_J$ is another parameter.

As stated above, the spin-fermion model with electron-phonon interactions 
will be studied with a Monte Carlo (MC) algorithm.
To simplify the numerical calculations, avoiding the sign problem, the
localized spins are assumed to be classical (with $\rm |S_{\bf i}|$=1).
This approximation is not drastic since most of the high Tc phenomenology 
is reproduced in this limit, and it was already discussed in 
detail in Ref. \cite{adri1}. 
Details of the MC method can be found in Ref.~\cite{yuno}. Square lattices 
with $8 \times 8$ and $12 \times 12$ sites will be studied here.

\section {Influence of Phonons on Striped States}

Neutron scattering experiments have shown that doping causes magnetic
incommensuration in the cuprates.\cite{Mason,cheong} The origin of this
phenomenon is still being debated. One possible scenario is the
formation of charge stripes in the ground state upon
doping.\cite{kivelson,zaanen} Experiments on nickelates such as
$\rm La_{2-x}Sr_xNiO_4$ (LSNO) have shown the
presence of diagonal static stripes \cite {Tranquada} and there is
evidence of stripes also in LNSCO, i.e., Nd doped ${\rm La_{2-x}Sr_xCu
O_4}$ (LSCO)\cite{tranquada2} and in ${\rm La_{2-x}Ba_xCuO_4}$
(LBCO).\cite{tranquada3} It is conjectured that the magnetic
incommensurability observed in other high $T_c$ cuprates is due to the
presence of dynamical stripes and that the stripe dynamics may be
related to the electron-phonon couplings in the different 
materials. Despite these theoretical scenarios, it has been very
difficult to find striped ground states in models for the cuprates
when studied with unbiased techniques. Stripes have been observed in the
$t-J$ model \cite{Steve,Elbio} but they are difficult to stabilize and
it is not clear whether the striped
state is the actual ground state or a very low lying excited
one. Stripes, on the other hand, have been obtained without biases in
the SF model that will be studied in this work.\cite{adri1}
This characteristic will allow us to explore the effect of
electron-phonon interactions in striped ground states. The charge
texture will be monitored by measuring the density structure factor
$N({\bf q})$.\cite{adri1} The magnetic structure factor $S({\bf q})$ will
provide information on the magnetic properties. Lattice distortions
will be monitored by measuring correlations between the displacements,
although the development of long range correlations in the lattice degrees
of freedom is not anticipated.

\begin{figure}[thbp]
\centerline{\psfig{figure=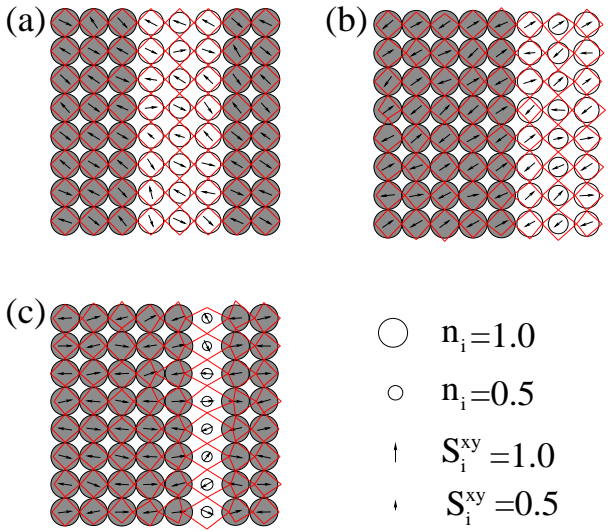,width=8cm}}
\vskip 0.3cm
\caption{(a) MC snapshot of an $8 \times 8$ lattice at 
$\langle n \rangle$=0.875 for 
$\lambda=0$ and $\gamma=0$. The size of the 
circles is proportional to the electronic density; 
the shaded circles have charge density larger than the average, i.e.,
$n_i\ge\langle n\rangle=0.875$. 
The arrows represent the projection of 
the localized spins in the plane $x-y$; the lines indicate 
lattice distortions (see text); (b) same as (a) but for $\lambda=1$ and mode 
$Q^{(2)}$; (c) same as (b) but for $\lambda=2$. }
\end{figure}

The dynamic properties of the system will be monitored by measuring the 
density of states $N(\omega)$, the one-particle 
spectral functions $A({\bf q},\omega)$ 
and the optical conductivity $\sigma(\omega)$.\cite{adri1,moham}

\subsection{Diagonal electron-phonon term:}


In our investigations it has been observed that in general 
the diagonal electron-phonon interaction plays an 
stabilizing role on stripe structures. This behavior was obtained for the four 
phonon modes studied here.
For $0\le\lambda\leq 2$ the holes 
become more localized in the stripes as $\lambda$ increases. 
This can be seen in Fig.1, where 
snapshots for $\langle n\rangle=0.875$  
are displayed for $\lambda=0$ (Fig.1a),
$\lambda=1$ (Fig.1b), and $\lambda=2$ (Fig.1c). The lines in the 
snapshots indicate the lattice distortions. If all the displacements 
$u_{{\bf i},\alpha}$ were 0 then, the lines would cross at the middle 
point of the links that join the lattice sites (as in Fig.1a). 
Out of center crossings 
indicate ionic displacements. It is clear from the figure that, as $\lambda$ 
increases, lattice distortions in the direction 
perpendicular to the stripe develop along the stripe with the mode $Q^{(2)}$, 
further localizing it.   
In Fig.1c large displacements along the horizontal direction can be seen in 
the links next to the stripe. The stripes become thinner 
and the density of holes per site inside the stripes increases. 

\begin{figure}[thbp]
\centerline{\psfig{figure=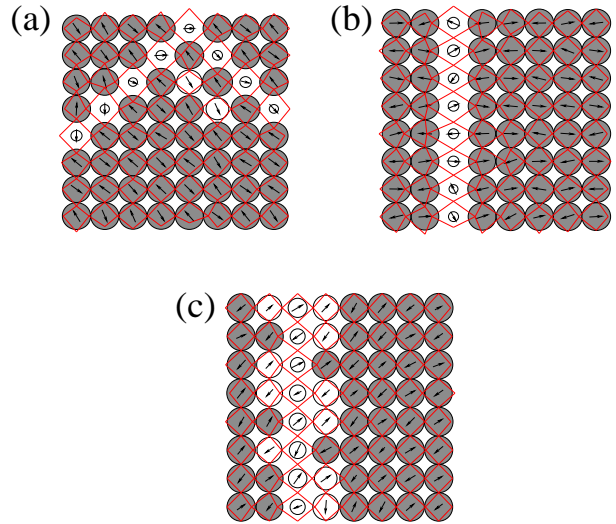,width=8cm}}
\vskip 0.3cm
\caption{(a) MC snapshot of an $8 \times 8$ lattice at 
$\langle n \rangle$=0.875, for mode $Q^{(1)}$, 
$\lambda=2$ and $\gamma=0$; (b) same as (a) but for mode $Q^{(2)}$;
(c) same as (a) but for mode $Q^{(3)}$. }
\end{figure}

We have observed some differences between the effects of the various phonon 
modes studied here as the strength of the diagonal 
electron-phonon coupling $\lambda$ increases. In Fig.2 snapshots 
for $\lambda=2$ at $\langle n \rangle=0.875$ 
are presented for $Q^{(1)}$, $Q^{(2)}$, 
and $Q^{(3)}$.\cite{footqy} A clear tendency to form {\it diagonal} 
stripes is seen 
for the breathing mode
$Q^{(1)}$ (Fig.2a). Since in this case 
holes are localized by being surrounded by 
four elongated bonds they cannot be accommodated in vertical or horizontal 
formations. Note that the single stripe that is stable at $\lambda=0$ 
for the electronic 
density shown in Fig.2 gets destabilized due to the $Q^{(1)}$ strong 
electron-phonon 
coupling. This result agrees with the fact that diagonal stripes are observed 
in LSNO
and experiments indicate that the breathing mode is the mode most strongly 
coupled to the electrons.\cite{trani} According to our results, 
a robust diagonal coupling of the electrons to the breathing mode should be 
expected in the nickelates.

In Fig.2b it can be observed that the shear mode $Q^{(2)}$ tends to 
stabilize vertical (or horizontal) stripes because a large horizontal (or
vertical) distortion occurs, localizing the holes along the stripe. 
Interestingly, the half-breathing mode also produces vertical (or horizontal)
stripes but the holes are less localized since the lattice can distort only 
along the horizontal (or vertical) direction. As a result, more dynamical 
stripes are observed for the half-breathing modes even for strong 
diagonal electron-phonon couplings (Fig.2c). Notice that the experimental
evidence indicates that in LNSCO, where vertical and horizontal stripes are 
observed, the mode more strongly coupled to the electrons is the 
half-breathing mode.\cite{Egami}

Another indication of increasing localization with  increasing $\lambda$
is observed in the peak in 
$N({\bf q})$, at ${\bf q}=(\pi/4,0)$ for the stripes shown in Fig.1a-c,
which becomes better developed as 
$\lambda$ increases (Fig.3a). In addition, the pseudogap in the density of 
states at 
the chemical potential becomes a full gap indicating an increase in insulating 
behavior (Fig.3b).

\begin{figure}[thbp]
\centerline{\psfig{figure=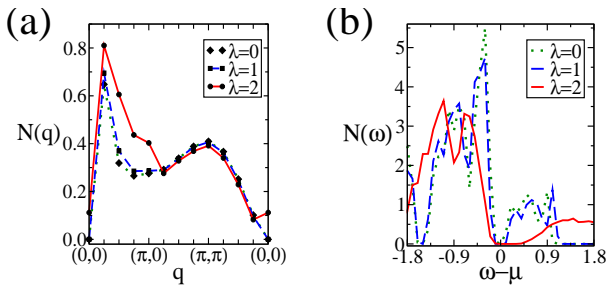,width=8cm}}
\vskip 0.3cm
\caption{(a) The charge structure factor $N({\bf q})$ for various values of 
$\lambda$ for the same parameters as in Fig.1;
(b) The density of states $N(\omega)$ for several values of the diagonal 
electron-phonon coupling $\lambda$, for the same parameters as in (a).The 
phonon mode is $Q^{(2)}$.}
\end{figure}

Similar results are obtained at $\langle n\rangle=0.75$ for which two 
stripes are stabilized, even at $\lambda=0$, in the $8\times 8$ systems 
studied here. 

The general trend, for different electronic densities and phonon 
modes, is that the ground state becomes more insulating as the diagonal 
electron-phonon coupling increases. The shear mode $Q^{(2)}$ will be 
used as an example but similar qualitative behavior is observed for the 
other modes. In Fig.4a it can be seen that the spectral weight at 
$\omega=0$ in the density of states decreases with increasing $\lambda$ 
for different values of the electronic density $\langle n\rangle$. The Drude 
weight, shown in Fig.4b, also decreases and insulating behavior is obtained 
for all densities at $\lambda=2$.

Although the effect of charge localization with increasing $\lambda$ is more 
pronounced at densities for which stripes are observed, we see 
in the snapshots
shown in Fig.5 for $\langle n \rangle=0.8$ and mode $Q^{(2)}$
that the charge becomes more localized as $\lambda$ increases 
and the lattice distortions localizing the holes develop large values 
for $\lambda \approx 2$. 
Note that AF domains separated by walls of holes are observed. These 
inhomogeneous structures appear to replace the stripes when the density of 
holes is not commensurate with the lattice size. An important characteristic 
of these 
inhomogeneous states is that, although no features are observed in the charge 
structure factor $N({\bf q})$, 
incommensurate magnetic correlations are still present and 
the peaks in $S({\bf q})$ occur 
at $(\pi,\pi-\delta)$ and $(\pi-\delta,\pi)$, i.e., 
in qualitative 
agreement with the incommensurate peaks observed in the cuprates.  

\begin{figure}[thbp]
\centerline{\psfig{figure=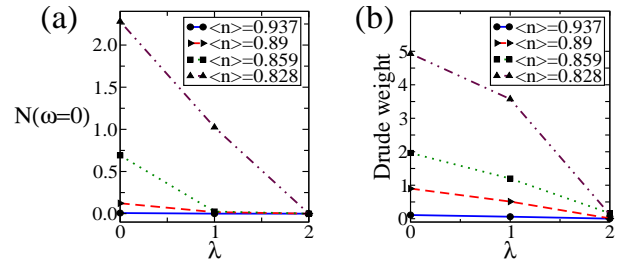,width=8cm}}
\vskip 0.3cm
\caption{(a) Spectral weight in the density of states $N(\omega)$ at 
$\omega=0$ as a function of the diagonal electron-phonon coupling $\lambda$
for several values of the electronic density $\langle n\rangle$ and mode 
$Q^{(2)}$; (b) Drude 
weight as a function of the diagonal electron-phonon coupling $\lambda$
for several values of the electronic density $\langle n\rangle$
and mode $Q^{(2)}$.}
\end{figure}

In 
the configurations shown in Fig.5, $S({\bf q})$ has a maximum at  
${\bf q}=(\pi,3\pi/4)$ for snapshot 
(a), but there is also a less intense peak at ${\bf q}=(3\pi/4,\pi)$. 

Incommensurate peaks in $S({\bf q})$ 
at ${\bf q}=(\pi,3\pi/4)$ and  ${\bf q}=(3\pi/4,\pi)$ with almost equal weight
are observed for $\langle n\rangle=0.8$ and $\lambda=1$ (see Fig.5d) although 
no peak in $N({\bf q})$ is observed. In
Fig.5b the snapshot of the 
final configuration of the corresponding Monte Carlo run is shown. 
Other snapshots of 
configurations appearing during the measuring part of our simulation are 
displayed in Fig. 6. It can be seen that the ground state is not frozen 
and that there is a dynamical charge redistribution. 
Thus, the magnetic 
incommensurability observed in some cuprates could be due to charge and spin 
configurations similar to those presented in Fig.6, which could be interpreted 
as ``dynamic stripes''. In fact, these results may indicate that
the patches in Fig.5 and 6 are not random islands since, 
if that were the case, we would expect that the maxima in $S({\bf q})$ would
form a ring in momentum space.\cite{crit} It is tempting to associate the 
observed patches with ``dynamical'' stripes.
Notice 
that for the states with ``static'' stripes only one peak at 
${\bf q}=(\pi,\pi-\delta)$ 
is observed in $S({\bf q})$ if the peak in 
$N({\bf q})$ is at ${\bf q}=(0,2\delta)$. On the other hand, the states with 
``dynamical'' stripes naturally reproduce the four peaks observed in 
neutron scattering experiments for the cuprates\cite{Mason,cheong} 
and the ``patch-like'' 
shape of the clusters would explain, at the same time, the apparently random 
inhomogeneous structures observed in STM experiments.\cite{Davis,Yazdani}

As the diagonal electron-phonon coupling $\lambda$ increases beyond 
$\lambda=2$,
important quantitative changes are observed, in particular, the tendency to 
bipolaron formation induced by strong lattice distortions. This behavior will 
be discussed in more detail in subsection C.

\begin{figure}[thbp]
\centerline{\psfig{figure=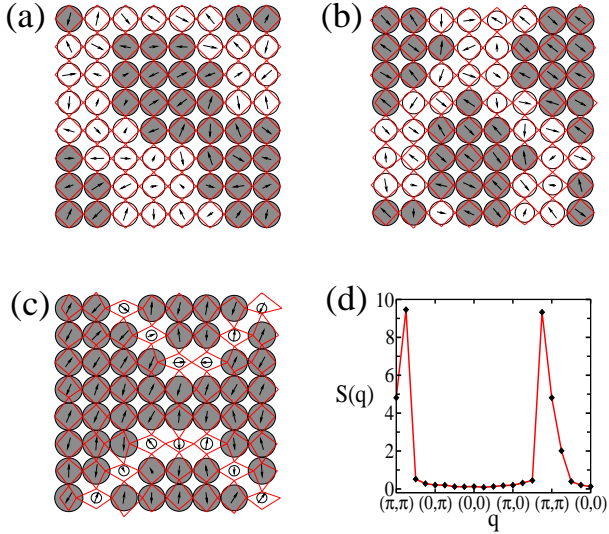,width=8cm}}
\vskip 0.3cm
\caption{(a) MC snapshot of an $8 \times 8$ lattice at 
$\langle n \rangle$=0.8 $\lambda=0$ and $\gamma=0$; (b) same as (a) but 
for $\lambda=1$ with mode $Q^{(2)}$;
(c) same as (b) for $\lambda=2$; (d) Magnetic structure factor for the 
parameters in (b).}
\end{figure}

\begin{figure}[thbp]
\centerline{\psfig{figure=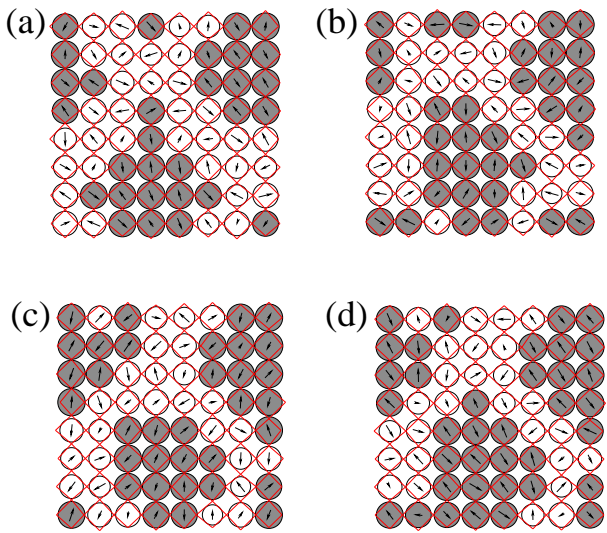,width=8cm}}
\vskip 0.3cm
\caption{(a) MC snapshot of an $8 \times 8$ lattice at 
$\langle n \rangle$=0.8, for mode $Q^{(2)}$, 
$\lambda=1$ and $\gamma=0$, after 2600 measuring sweeps; (b) same as 
(a) but after 3750 measuring sweeps;
(c) same as (a) but after 4250 measuring sweeps;
(d) same as (a) but after 5000 measuring sweeps.}                    
\end{figure}

\subsection{Off-Diagonal electron-phonon term:}

\begin{figure}[thbp]
\centerline{\psfig{figure=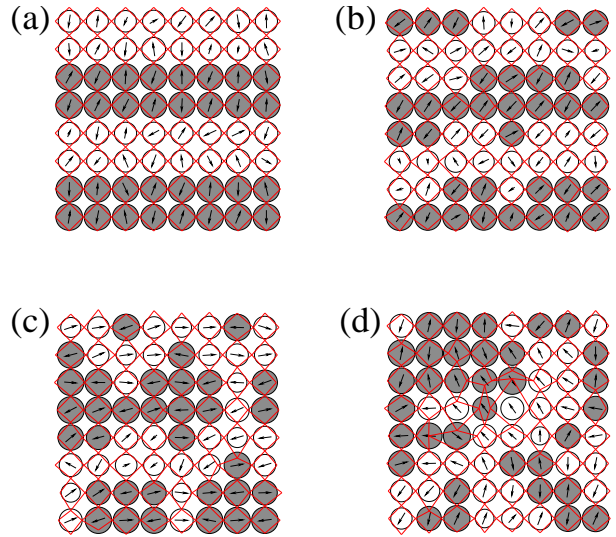,width=8cm}}
\vskip 0.3cm
\caption{Study of the effect of off-diagonal couplings. (a) MC 
snapshot of an $8 \times 8$ lattice at 
$\langle n \rangle$=0.75,
$\lambda=0$ and $\gamma=0$; (b) same as (a) but for $\gamma=0.1$ and 
mode $Q^{(2)}$;
(c) same as (b) for $\gamma=0.2$; (d) same as (b) but for $\gamma=0.6$ . }
\end{figure}

\begin{figure}[thbp]
\centerline{\psfig{figure=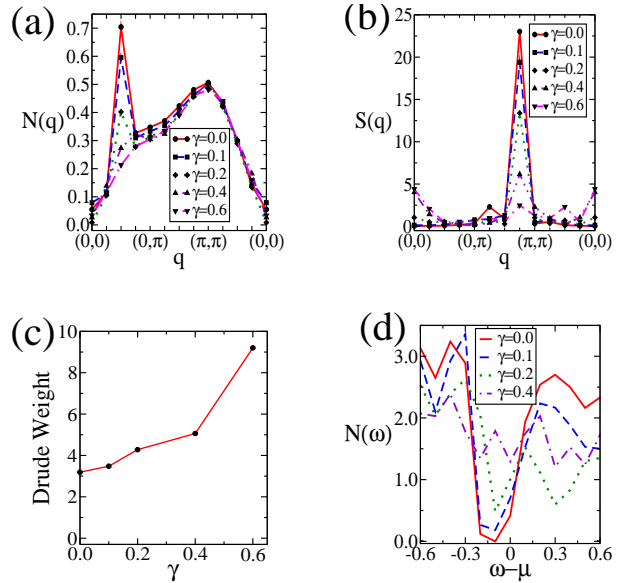,width=8cm}}
\vskip 0.3cm
\caption{(a) Charge structure factor for different values of $\gamma$ (strength
of the off-diagonal EPI) and 
for $\lambda=0$ on a $8 \times 8$ lattice at 
$\langle n \rangle$=0.75, for mode $Q^{(2)}$; 
(b) the magnetic structure factor for the same parameters as in (a);
(c) the Drude weight for the same parameters as in (a);
(d) the density of states for the same parameters as in (a). }
\end{figure}

In this first exploratory study of the effects of the off-diagonal terms 
due to the EPI we will allow the parameter 
$\gamma$ in Eq.(8) and Eq.(10) 
to vary in the interval $(0,0.6)$, while $g_J$ (see Eq.(10)) will be 
kept equal to 1. In general, we have observed that the effect of the 
off-diagonal term is to destabilize the stripes, since they 
become more dynamic.
Examples of this effect can be seen in the snapshots presented in Fig.7 for
$\langle n \rangle=0.75$ and $\gamma=0$, 0.1, 0.2 and 0.6. The stripes 
become distorted as $\gamma$ increases and, eventually, AF domains separated 
by irregularly shaped 
hole-rich regions start to develop.

In Fig.8a, it is shown how the sharp 
maximum in the charge structure factor loses intensity as the stripes become 
more dynamic. However, notice that the magnetic structure factor (Fig.8b) 
still shows incommensurability at ${\bf q}=(\pi,3\pi/4)$ which is in agreement 
with neutron scattering data for dynamic stripes. For $\gamma\ge 0.4$ a peak 
in $S({\bf q})$ develops at ${\bf q}=(0,0)$. 
This seems to occur because the hole 
domains become ferromagnetic as it can be seen in Fig.7c and d.

The system 
also becomes more metallic since the Drude weight increases with $\gamma$
(Fig.8c) and more spectral weight appears in the pseudogap in the DOS 
(Fig.8d). 

\subsection{Diagonal and Off-Diagonal terms:}

When both diagonal and non-diagonal electron-phonon couplings are active
simultaneously, we have observed that for values of $\lambda\leq 2$, 
there is a competition between the localizing effect of the diagonal
electron-phonon coupling $\lambda$ and the disordering tendency of the 
off-diagonal parameter $\gamma$. As a result, larger values of $\gamma$ than 
in the case of $\lambda=0$ are 
needed to destabilize the stripes. The stripe states are replaced by AF 
``clusters'' separated by hole-rich regions as shown in Fig.9.

\begin{figure}[thbp]
\centerline{\psfig{figure=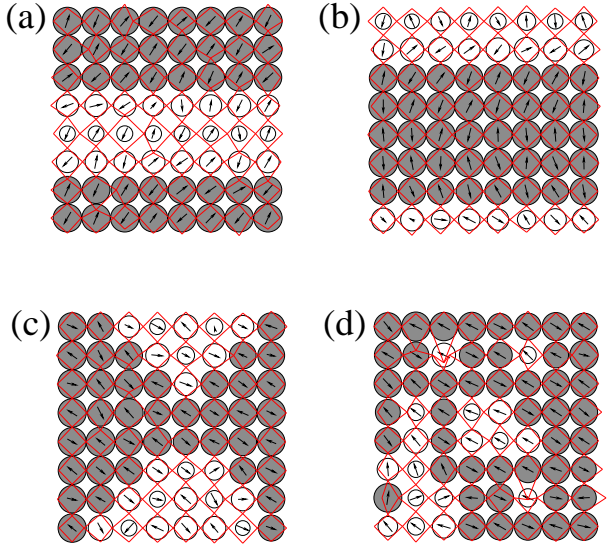,width=8cm}}
\vskip 0.3cm
\caption{(a) MC snapshot of an $8 \times 8$ lattice at 
$\langle n \rangle$=0.875, for mode $Q^{(2)}$, 
$\lambda=1$ and $\gamma=0$; (b) same as (a) but for $\gamma=0.1$;
(c) same as (a) but for $\gamma=0.2$;
(d) same as (a) but for $\gamma=0.4$. }
\end{figure}

We also have observed that for $\lambda\ge 3$ the modes $Q^{(1)}$ and 
$Q^{(2)}$ induce charge density wave (CDW) states. CDW domains are formed
in order to accommodate the extra holes away from half-filling as shown 
in Fig.10a for $\langle n\rangle=0.875$ and mode $Q^{(2)}$. This is the 
only case in which we have observed long range order developing in the 
lattice degrees of freedom. The dashed lines in the figure indicate 
different CDW domains. In Fig.10b it can be seen how the 
off-diagonal coupling destabilizes the CDW state and a disordered state with 
bipolarons (two electrons trapped at the same site) is observed. 

As $\gamma$ increases the off-diagonal term opposes the trend towards 
bipolaron formation caused by the diagonal electron-phonon coupling and 
stripe-like structures reappear for some dopings. 

\begin{figure}[thbp]
\centerline{\psfig{figure=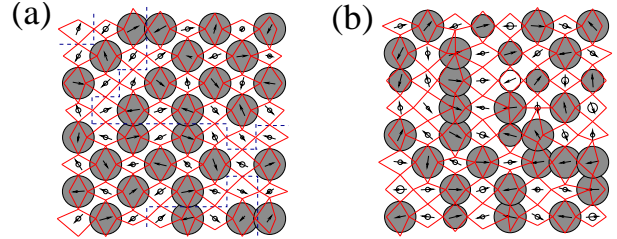,width=8cm}}
\vskip 0.3cm
\caption{(a) MC snapshot of an $8 \times 8$ lattice at 
$\langle n \rangle$=0.875, for mode $Q^{(2)}$, 
$\lambda=6$ and $\gamma=0$. The dashed lines separate CDW domains; 
(b) same as (a) but for $\gamma=0.2$.}
\end{figure}

The half-breathing modes, on the other hand, are not able to stabilize a CDW 
state even for large values of diagonal couplings and they only produce 
disordered states with bipolarons when the diagonal electron-phonons 
coupling is large. Since CDW and superconducting states normally compete with 
each other, the half-breathing mode may enhance the development of 
superconductivity. Unfortunately, off-diagonal long-range order which 
characterizes S or D wave superconductivity, cannot develop with 
adiabatic phonons.\cite{raedt} Thus, this possibility should be explored 
for finite values of the frequency of the lattice vibrations in a full 
quantum calculation.

\subsection{Spectral Functions:}

In this subsection the properties of the spectral functions 
$A({\bf q},\omega)$ will 
be discussed in detail. 

Numerical studies of the spin-fermion model 
without phonons in Ref.\cite{moham} have 
shown that the underdoped regime is characterized by a depletion of spectral 
weight along the diagonal direction in momentum space, where a very weak Fermi 
surface (FS) may exist. On the other hand,
strong spectral weight, very flat bands, and a well defined Fermi surface (FS)
are observed close to 
${\bf q}=(\pi,0)$ and $(0,\pi)$. 
These results are in agreement with ARPES 
studies for LSCO,\cite{Ino} material believed to have dynamic stripes, and also 
with numerical studies of models in which stripes have been built via a 
configuration dependent ``stripe'' potential in the t-J model\cite{Sada}
or with stripe-like mean-field states.\cite{Oles} ARPES results for LNSCO, 
where stripes have been observed,\cite{tranquada2} indicate that the 
low-energy excitation near the expected d-wave node region is strongly 
suppressed.\cite{zhou}

In the absence of EPI the above mentioned characteristics are well 
reproduced by the SF model.
In Fig.11a the spectral function $A({\bf q},\omega)$ along the path 
${\bf q}=(0,0)-(\pi,0)-$ 
$(\pi,\pi)$ for $\langle n \rangle=0.80$ on a $12 \times 12$ lattice
in the absence 
of EPI is presented. A well defined quasi-particle peak is observed at $(0,0)$,
a flat band appears close to 
$(\pi,0)$ and the peak crosses the Fermi energy at 
$(\pi,5\pi/6)$ defining a hole-like FS. The spectral function along the 
diagonal of the Brillouin zone is presented in Fig.11b. It can be seen how 
the spectral weight becomes very incoherent as the Fermi energy is reached.

\begin{figure}[thbp]
\centerline{\psfig{figure=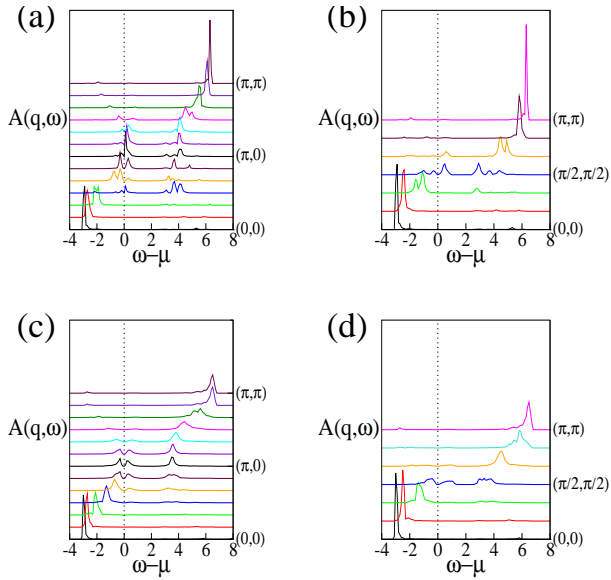,width=8cm}}
\vskip 0.3cm
\caption{(a) Spectral functions along the path 
${\bf q}=(0,0)-$ $(\pi,0)-(\pi,\pi)$ on a $12 \times 12$ lattice at 
$\langle n \rangle$=0.8,
$\lambda=0$ and $\gamma=0$; (b) same as (a) but along the diagonal 
direction in the Brillouin zone; (c) same as (a) but for $\lambda=2$ and mode
$Q^{(3)}$; (d) same as (b) but for $\lambda=2$.}
\end{figure}

\begin{figure}[thbp]
\centerline{\psfig{figure=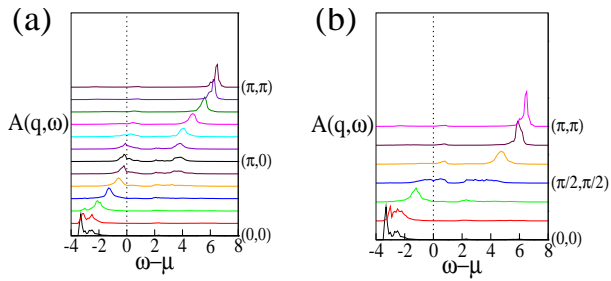,width=8cm}}
\vskip 0.3cm
\caption{(a) Spectral functions along the path 
${\bf q}=(0,0)-$ $(\pi,0)-(\pi,\pi)$ on a $12 \times 12$ lattice at 
$\langle n \rangle$=0.8,
$\lambda=2$ and $\gamma=0.4$ for the half-breathing mode; 
(b) same as (a) but along the diagonal 
direction in the Brillouin zone.}
\end{figure}
As a general rule we have observed that electron-phonon interactions increase 
the decoherence of the spectral functions, particularly close to the Fermi 
energy. Below, we will discuss the effects of the different phonon 
modes for the most interesting case of dynamic stripes 
(see, for example, Fig.6).  


The diagonal electron-phonon coupling renders the system insulator, as 
discussed in subsection A. In Fig.11c the spectral weight along the path
${\bf q}=(0,0)-(\pi,0)-(\pi,\pi)$ is displayed for $\lambda=2$ and the 
half-breathing mode $Q^{(3)}$ in a $12 \times 12$ lattice. 
It can be seen that the quasiparticles are less well defined and that a 
clear gap has opened at the Fermi energy. Along the diagonal of the Brillouin 
zone (Fig.11d) the spectral weight becomes more incoherent and a gap also is 
observed.

The off-diagonal EPI restores spectral weight in the gap but there are no 
quasiparticles close to the Fermi energy as it can be seen in Fig.12a and b
where the spectral functions for $\lambda=2$, $\gamma=0.4$  and mode 
$Q^{(3)}$ are shown along
${\bf q}=(0,0)-(\pi,0)-(\pi,\pi)$ and along the diagonal of the Brillouin zone.
For $\gamma$ finite and $\lambda=0$ the only observed changes is that the peaks
in Figs.11a and b become broader.


For $\lambda=2$ the breathing mode 
creates a 
larger insulating gap than the half-breathing mode and produces a larger 
decoherence of the quasiparticle peaks. The results for the shear mode are 
qualitatively very similar.



The off-diagonal EPI for the shear and breathing modes tends to close the 
gap and increase the decoherence.

\section{Influence of Phonons on Homogeneous States and generation of Stripes}

Up to this point we have considered the effects of EPI on ground states that 
already presented charge inhomogeneity. However, it is important to study 
whether the electron-phonon interactions proposed in this work can themselves
generate charge inhomogeneity, in particular stripes, in a previously 
homogeneous ground state.

In order to address this issue, we studied the S-F model with $J=1.5$ 
instead of $J=2$, value which was used in the previous sections (all the other 
parameters are kept the same). For $\langle n \rangle =0.75$ the ground state 
has an homogeneous charge distribution as it can be observed 
in the MC snapshot 
presented in Fig.13a. Note that despite the charge homogeneity 
this state presents magnetic incommensurability due to a spiral spin 
arrangement in the vertical direction. 

One of the main results of this paper is our observation 
that a strong diagonal coupling with the shear mode {\it generates}
two horizontal or vertical stripes. An example can be seen in the snapshot 
presented in Fig.13b for $\lambda=2$. In this case 
the holes act as boundaries between
undoped antiferromagnetic states and the magnetic incommensurability arises 
from the inhomogeneous charge distribution. A $\pi$-shift among the AF domains 
is observed as well.\cite{hb}

The breathing mode also induces charge 
inhomogeneity for a diagonal coupling $\lambda=2$. From the discussion in 
Section III-A diagonal stripes would be expected but we have observed
two stripes with a zig-zag shape, i.e, the holes align diagonally at 
short distance scales but, on the whole, the stripe is still 
horizontal or vertical (see Fig.13c). 

\begin{figure}[thbp]
\centerline{\psfig{figure=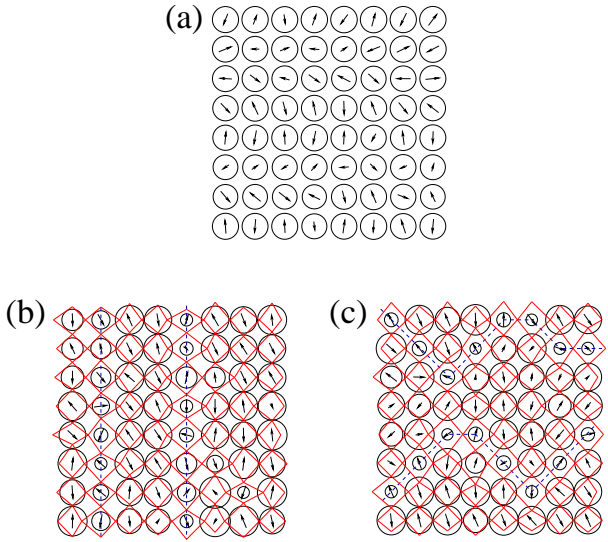,width=8cm}}
\vskip 0.3cm
\caption{(a) MC snapshot of an $8 \times 8$ lattice at 
$\langle n \rangle$=0.75, $J=1.5$,
$\lambda=0$ and $\gamma=0$; (b) same as (a) but for $\lambda=2$ with mode 
$Q^{(2)}$;
(c) same as (b) with mode $Q^{(1)}$ . The dashed line indicates the stripes.}
\end{figure}

This result indicates that {\it diagonal EPI are able to induce stripe-like 
charge inhomogeneities in otherwise homogeneous states}. However, 
the S-F model also
shows that although diagonal EPI stabilize stripes they are 
not {\it necessary} to induce them. Charge inhomogeneities can result even 
in the absence of EPI, just from magnetic interactions. 
  
\section{Conclusions}

Summarizing, we have studied the effects of diagonal and off-diagonal 
electron-phonon interactions using non-biased numerical techniques on a model 
that has both striped and homogeneous ground states in the absence of EPI. 
We found that diagonal EPI tend to either generate or further
stabilize the existing 
stripes and turn the system into an insulator. Horizontal and vertical 
stripes are stabilized by half-breathing and shear modes, but the stripes 
generated by shear modes are more localized. The 
breathing mode stabilizes static diagonal stripes for intermediate 
diagonal couplings. Large diagonal couplings stabilize CDW states for 
breathing and shear modes but this state is not observed with the 
half-breathing mode. 

On the other hand, off-diagonal electron-phonon couplings  
destabilize the stripes making the ground state more metallic, 
although non homogeneous. Instead of static stripes other kinds of 
inhomogeneous structures 
characterized by antiferromagnetic domains separated by barriers of holes are 
observed. 

The electron-phonon interactions undermine the quasiparticle peaks in the 
spectral functions close to the Fermi energy producing incoherent weight.
Breathing and shear modes tend to open large gaps at the Fermi energy
creating insulating behavior. The half-breathing mode, on the other hand, 
opens smaller gaps that are closed by relatively modest off-diagonal 
couplings,
allowing for metallic behavior.

Our results indicate that the half-breathing mode is most likely to play a 
role in non-insulating  materials with vertical and/or horizontal ``dynamic'' 
stripes, such as LSCO, 
while the breathing mode should dominate on insulators with 
diagonal stripes like the nickelates.    

The most important result of this study is the fact that different phonon 
modes promote a diverse array of charge structures and that the relative 
strength of diagonal and off-diagonal couplings influences the transport 
properties. Diagonal electron-phonon couplings promote insulating behavior,
while off-diagonal interactions are crucial to achieve metalicity. It appears
that the half-breathing mode off-diagonally coupled to the electrons is the
most likely to  produce non-insulating states with dynamical stripes as
observed in the cuprates. This is in agreement with the experimental data 
which indicate the prevalence of half-breathing modes in the high Tc 
cuprates.\cite{Egami} 
The electron-phonon interaction introduces decoherence of the 
quasi-particle peak in the spectral function in agreement with ARPES
measurements in LSCO. 

The crucial issue that remains 
to be explored is whether electron-phonon interactions are
needed, in addition to the magnetic exchange, in order to develop long range 
D-wave pairing correlations. Since superconductivity arises from off-diagonal 
long range order, it cannot be generated with adiabatic phonons. The next step
will be to study off-diagonally coupled half-breathing modes at finite 
frequencies.

\section{Acknowledgments}

We acknowledge discussions with T. Egami, J. Tranquada, G. Sawatsky and E. 
Dagotto. A.M. is supported by NSF under grants DMR-0443144 and DMR-0454504.
Additional support is provided by ORNL.


\begin{references}

\bibitem{adri1} C.~Buhler, S. Yunoki and A. Moreo, 
Phys.~Rev.~Lett.~{\bf 84}, 2690 (2000); C. Buhler, S. Yunoki and A. Moreo,
Phys. Rev. B{\bf 62}, R3620 (2000).


\bibitem{BCS} J.Bardeen, L.N. Cooper and J.R. Schrieffer, Phys. Rev. {\bf 108}, 
1175 (1957).

\bibitem{magnetic} P.W. Anderson and J.R. Schrieffer, Phys. Today {\bf 44} 
55 (1991).

\bibitem{review}E. Dagotto, Rev. Mod. Phys.{\bf 66}, 763 (1994).

\bibitem{Bianconi} A. Bianconi, N.L. Saini, A. Lanzara, M. Missori, 
T. Rossetti, H. Oyanagi, H. Yamaguchi, K.Oka and T. Ito, Phys. Rev. Lett. 
{\bf 76} 3412 (1996).

\bibitem{Egami} R.J. McQueeney, Y. Petrov, T. Egami, M.Yethiraj, G, Shirane 
and Y. Endoh, Phys.Rev.Lett.{\bf 82}, 628 (1999).

\bibitem{Lanzara} A. Lanzara, P. V. Bogdanov, X. J. Zhou, S. A. Kellar, 
D. L. Feng, E. D. Lu, S. Uchida, H. Eisaki, A. Fujimori, K. Kishio, 
J.-I. Shimoyama, T. Noda, S. Uchida, Z. Hussain, and Z.-X. Shen, 
Nature {\bf 412}, 510 (2001).

\bibitem{Tranquada} J.M. Tranquada, D.J. Buttrey, V. Sachan and J.E. Lorenzo, 
Phys.Rev.Lett.{\bf 73}, 1003 (1994); J.M. Tranquada, R. Mallozi, J. Orenstein,
T.N. Eckstein and I. Bozovic, 
Nature (London)) {\bf 375}, 561 (1995).

\bibitem{Davis} K. McElroy, R.W. Simmonds, J.E. Hoffman, D.-H. Lee, 
J. Orenstein, H. Eisaki, S. Uchida, and J.C. Davis, 
Nature {\bf 422}, 520 (2003).

\bibitem{Yazdani} M. Vershinin, S. Misra, S. Ono, Y. Abe, Y. Ando, and A. 
Yazdani, www.sciencexpress.org, 10.1126/science.1093384. 

\bibitem{manrev} E. Dagotto, T. Hotta and A. Moreo, Phys. Rep. {\bf 344}, 1 
(2001).

\bibitem{bili} T. Egami and S. Billinge, Physical Properties of High 
Temperature Superconductors V, ed. D. M. Ginsberg 
(World Scientific, Singapore), page 265 (1996).

\bibitem{Olle} O.~R\"osch and O.~Gunnarsson, Phys. Rev. Lett. {\bf 92}, 
146403 (2004).

\bibitem{Dev} T.P. Devereaux, T. Cuk, Z-X. Shen and N. Nagaosa, preprint. 
Cond-mat/0403766.

\bibitem{jose} A. Dobry, A. Greco, S. Koval and J. Riera, Phys.~Rev.~B
{\bf 52}, 13722 (1995).

\bibitem{doug} T. Sakai, D. Poilblanc and D.J. Scalapino, Phys.~Rev.~B
{\bf 55}, 8445 (1997).

\bibitem{bishop} K. Yonemitsu, A.R. Bishop and J. Lorenzana, Phys.~Rev.~B
{\bf 47}, 8065 (1993).

\bibitem{ishihara} S. Ishihara and N. Nagaosa, preprint. Cond-mat/0311200. 

\bibitem{tachi} S. Ishihara, T. Egami, and M. Tachiki, Phys.~Rev.~B
{\bf 55}, 3163 (1997); Y. Petrov and T. Egami, Phys.~Rev.~B
{\bf 58}, 9485 (1998).

\bibitem{Pines} A.~Chubukov, Phys.~Rev.~B{\bf 52}, R3840 (1995);
S.~Klee and A.~Muramatsu, Nucl. Phys. {\bf B 473}, 539 (1996);
J.R.~Schrieffer, 
J. of Low Temp.~Phys.~{\bf 99}, 397 (1995);
B.L.~Altshuler, L.B. Yoffe and A.J. Millis,
Phys.~Rev.~B{\bf 52}, 5563
(1995); C.-X.~Chen and H.-B.~Schüttler,
Phys.~Rev.~B{\bf 43}, 3771 (1991).

\bibitem{moham} M. Moraghebi, C. Buhler, S. Yunoki and
A. Moreo. Phys. Rev. B{\bf 63}, 214513 (2001).

\bibitem{yuno} E.~Dagotto, , S. Yunoki, A.L. Malvezzi, A. Moreo, J. Hu, 
S. Capponi and D. Poilblanc, Phys. Rev. B{\bf 58}, 6414 (1998).

\bibitem{Mason} T.E. Mason, G. Aeppli and H.A. Mook,  
Phys.Rev.Lett.{\bf 68}, 1414 (1992).

\bibitem{cheong} S.-W. Cheong, G. Aeppli, T. E. Mason, H. Mook, 
S. M. Hayden, P. C. Canfield, Z. Fisk, K. N. Clausen, and J. L. Martinez,
Phys.Rev.Lett.{\bf 67}, 1791 (1991); P.Dai, H. A. Mook, and F. Dogan
Phys.Rev.Lett.{\bf 80}, 1738 (1998) . 

\bibitem{kivelson} O. Zachar, S. Kivelson and V. Emery,
Phys.~Rev.~B{\bf 57}, 1422 (1998). 

\bibitem{zaanen} J. Zaanen, Science {\bf 286} 251 (1999).

\bibitem{tranquada2} J.M. Tranquada, J.D. Axe, N. Ichikawa, Y. Nakamura, 
S. Uchida and B. Nachumi, Phys. Rev. {\bf B 54}, 7489 (1996); 
J.M. Tranquada, J.D. Axe, N. Ichikawa, A.R. Moodenbaugh, Y. Nakamura and
S. Uchida, Phys.Rev.Lett.{\bf 78}, 338 (1997).

\bibitem{tranquada3} M. Fujita, H. Goka, K. Yamada, J.M. Tranquada and 
L.P. Regnault, Phys. Rev. {\bf B 70}, 104517 (2004).

\bibitem{Steve} S.R. White and D.J. Scalapino, 
Phys.Rev.Lett.{\bf 81}, 3227 (1998).

\bibitem{Elbio} G.B. Martins, R. Eder and E. Dagotto, Phys. Rev. {\bf B 60}, 
R3716 (1999).

\bibitem{footqy} The only difference between $Q^{(4)}$ and $Q^{(3)}$, is that
$Q^{(4)}$ encourages horizontal stripes instead of vertical ones like 
$Q^{(3)}$.            

\bibitem{trani} J.M. Tranquada, K.Nakajima, M. Braden, L. Pintschovius and R.
McQueeney, Phys.Rev.Lett.{\bf 88}, 075505 (2002).

\bibitem{crit} Since we work in finite lattices there is a discrete number of 
points in momentum space and additional peaks in $S({\bf q})$ at values of 
${\bf q}$ not sampled in our lattices cannot be ruled out. 

\bibitem{raedt} L.A. van Dijk, Ph.D. Thesis, University of Groningen, 1998.

\bibitem{Ino} A. Ino, C. Kim, M. Nakamura, T. Yoshida, T. Mizokawa, 
A. Fujimore, Z.-X. Shen, T. Kakeshita, H. Eisaki, and S. Uchida,
 Phys. Rev. B{\bf 65}, 094504 (2002).

\bibitem{Sada} T. Tohyama, S. Nagaki, Y. Shibata, and S. Maekawa, 
Phys.Rev.Lett.{\bf 82}, 4910 (1999).

\bibitem{Oles} M. Fleck, A. Lichtenstein, and E. Pavarini,
Phys.Rev.Lett.{\bf 84}, 4962 (2000).

\bibitem{zhou} X.J. Zhou, P. Bogdanov, S.A. Kellar, T. Noda, H. Eisaki, 
S. Uchida, Z. Hussain and Z.-X. Shen, Science {\bf 286}, 268 (1999).

\bibitem{hb} We have observed that the half-breathing mode induces 
charge inhomogeneity but not stripes.

\end{references}
\end{document}